\documentclass[prb,onecolumn,preprintnumbers,amsmath,amssymb,floatfix,showpacs]{revtex4}

\usepackage[dvips]{graphics,graphicx,color}

\begin{document}

\title{Quantum Monte Carlo study of the ground state of the
  two-dimensional Fermi fluid}

\author{N.\ D.\ Drummond and R.\ J.\ Needs}

\affiliation{TCM Group, Cavendish Laboratory, University of Cambridge,
J.\ J.\ Thomson Avenue, Cambridge CB3 0HE, United Kingdom}

\begin{abstract}
We have used the variational and diffusion quantum Monte Carlo methods
to calculate the energy, pair correlation function, static structure
factor, and momentum density of the ground state of the
two-dimensional homogeneous electron gas.  We have used highly
accurate Slater-Jastrow-backflow trial wave functions and twist
averaging to reduce finite-size effects where applicable.  We compare
our results with others in the literature and construct a
local-density-approximation exchange-correlation functional for 2D
systems.
\end{abstract}


\maketitle

\section{Introduction \label{sec:introduction}}

The homogeneous electron gas (HEG) plays a key role in modern
condensed-matter theory.  It consists of a set of electrons moving in
a uniform, inert, neutralizing background, and is the simplest
fully-interacting quantum many-body model of condensed matter.  The
three-dimensional HEG models the behavior of the conduction electrons
in metals and semiconductors, while the two-dimensional (2D) HEG
models the behavior of electrons confined to surfaces and thin layers.
Quantum Monte Carlo (QMC) methods\cite{ceperley_1980,foulkes_2001}
have long played an important role in establishing the ground-state
properties of the HEG\@.  Most effort has gone into calculating the
ground-state energy of different phases as a function of density, in
order to establish the zero-temperature phase
diagram.\cite{tanatar,rapisarda,ndd_2Dheg} In this article, we report
QMC calculations of some other properties of the 2D HEG of interest to
condensed-matter physicists: the pair correlation function (PCF),
static structure factor (SSF), and momentum density (MD)\@.  We also
report energy data for high-density HEG's. We have confined our
attention to the fluid phase, which is the ground state at the
densities typically encountered in experiments.

The PCF, especially the contact PCF $g(0)$, is a key ingredient in
generalized-gradient-approximation exchange-correlation functionals
for density functional theory (DFT) calculations.  The PCF has been
studied several times using
QMC,\cite{tanatar,rapisarda,kwon_pcf,gorigiorgi_pcf} but the value of
$g(0)$ at low densities has proved controversial because electrons
approach one another infrequently, and the QMC results disagree with
values calculated using ladder theory.\cite{qian} Our QMC data,
obtained using a different trial wave function from the earlier
calculations, should help to clarify the situation.  The SSF is
related to the PCF by a Fourier transform.  SSF data at small $k$ are
needed to establish the long-range behavior of the PCF\@.

The MD of the HEG is of considerable importance in Fermi liquid
theory. To our knowledge, the only QMC MD data to have been published
for the 2D HEG are those of Tanatar and Ceperley,\cite{tanatar} which
used a relatively simple form of trial wave function.  (A fit to QMC
data generated by Conti\cite{conti} is reported in Ref.\
\onlinecite{giuliani}, but no details about the calculations are
given.) Tanatar and Ceperley's low-density MD shows a very strange
feature: the MD is lower at zero momentum than it is at the top of the
Fermi edge.  It is clearly important to provide new QMC MD data, in
order to establish whether this is a genuine property of the HEG\@.

Finally, we report ground-state energy data for paramagnetic Fermi
fluids, which we use to parameterize a local-density-approximation
exchange-correlation functional for use in DFT studies of 2D systems.

The rest of this article is arranged as follows.  In Sec.\
\ref{sec:qmc_calcs} we describe the computational techniques used.  In
Sec.\ \ref{sec:results} we present the data we have generated.
Finally, we draw our conclusions in Sec.\ \ref{sec:conclusions}.
Densities are given in terms of the radius $r_s$ of the circle that
contains one electron on average.  We use Hartree atomic units
($\hbar=|e|=m_e=4\pi\epsilon_0=1$) throughout this article.  All our
QMC calculations were performed using the \textsc{casino}
code.\cite{casino}

\section{QMC calculations \label{sec:qmc_calcs}}

\subsection{Trial wave functions \label{sec:trial_wfs}}

In the variational quantum Monte Carlo (VMC) method, expectation
values are calculated with respect to an approximate trial wave
function, the integrals being performed by a Monte Carlo technique.
In diffusion quantum Monte Carlo\cite{ceperley_1980,foulkes_2001}
(DMC) the imaginary-time Schr\"odinger equation is used to evolve an
ensemble of electronic configurations towards the ground state.  The
fermionic symmetry is maintained by the fixed-node
approximation,\cite{anderson_1976} in which the nodal surface of the
wave function is constrained to equal that of a trial wave function.
The VMC algorithm generates electron configurations distributed
according to the square of the trial wave function, while the DMC
algorithm generates configurations distributed as the product of the
trial wave function and its ground-state component.

Our trial wave functions consisted of Slater determinants of
plane-wave orbitals multiplied by a Jastrow correlation factor. The
Jastrow factor contained polynomial and plane-wave expansions in
electron-electron separation.\cite{ndd_jastrow} The orbitals in the
Slater wave function were evaluated at quasiparticle coordinates
related to the actual electron positions by backflow functions
consisting of polynomial expansions in electron-electron
separation.\cite{backflow} The wave functions were optimized by
variance minimization\cite{umrigar_1988a,ndd_newopt} and
linear-least-squares energy minimization.\cite{umrigar_emin}

We simulated HEG's in finite, square cells subject to periodic
boundary conditions.  The many-body Bloch theorem\cite{rajagopal}
states that the wave function $\Psi$ satisfies
\begin{equation} \Psi({\bf r}_1,\ldots,{\bf r}_i+{\bf
R}_s,\ldots,{\bf r}_N)=\exp(i{\bf k}_s \cdot {\bf R}_s) \Psi({\bf
    r}_1,\ldots,{\bf r}_N), \end{equation} where ${\bf R}_s$ is a
    simulation-cell lattice vector and ${\bf k}_s$ is the
    simulation-cell Bloch vector.  In some of our calculations, and in
    previous QMC studies of the 2D HEG,\cite{tanatar,rapisarda,kwon}
    it has been assumed that ${\bf k}_s={\bf 0}$.  However, in our
    calculations of the energy, PCF, and SSF we performed twist
    averaging, in which expectation values are averaged over ${\bf
    k}_s$ in the first Brillouin zone of the simulation
    cell.\cite{lin_twist_av} This procedure greatly reduces
    single-particle finite-size errors caused by shell-filling effects.

The high quality of our trial wave functions is demonstrated in Table
\ref{table:wf_quality_rs5}, which shows QMC energies achieved using
different levels of wave function for a 58-electron paramagnetic Fermi
fluid of density parameter $r_s=5$ a.u.  Backflow functions change the
nodal surface of the trial wave function and can therefore improve the
fixed-node DMC energy.  In practice we find that backflow lowers the
DMC energy substantially.  Our VMC energies are significantly lower
than those of Kwon \textit{et al.},\cite{kwon} as is our
Slater-Jastrow-backflow DMC energy.  On the other hand, the
Slater-Jastrow-backflow DMC energy of Attaccalite \textit{et
al.}\cite{attaccalite}\ is higher than that of Kwon \textit{et al.}
Our Slater-Jastrow DMC energy is slightly lower than that of
Attaccalite \textit{et al.}, which in turn is lower than that of Kwon
\textit{et al.}  Since the nodal surface is the same in the three
calculations, these DMC energies really ought to agree.  However, the
trial wave function used by Kwon \textit{et al.}\ is very much poorer
than ours, as can be seen by comparing the VMC energies.  Time-step
and population-control biases in their DMC energies must be much
greater, which may explain the discrepancy.  The results of
Attaccalite \textit{et al.}\ have not been extrapolated to zero time
step, hindering comparison. The VMC and DMC results of Rapisarda and
Senatore\cite{rapisarda} are in very close agreement with those of
Kwon \textit{et al.}\cite{kwon}

\begin{table}
\begin{center}
\begin{tabular}{lr@{.}lr@{.}lr@{}l}
\hline \hline

Method & \multicolumn{2}{c}{Energy (a.u.\ / elec.)} &
\multicolumn{2}{c}{Var.\ (a.u.)} & \multicolumn{2}{c}{Frac.\ corr.\
en.} \\

\hline

HF             & ~~~$-0$&$100\,222\,006$ &
\multicolumn{2}{c}{$\cdots$} & 0&\% \\

SJ-VMC         & $-0$&$148\,211\,0(8)$ & ~~$0$&$019\,6$ &
~~~96&.910(4)\% \\

SJ-VMC$^\ast$  & $-0$&$146\,80(5)$     & \multicolumn{2}{c}{$\cdots$}
& 94&.1(1)\%  \\

SJB-VMC & $-0$&$149\,385\,1(6)$ & $0$&$007\,74$ & 99&.282(4)\% \\

SJB-VMC$^\ast$ & $-0$&$148\,80(5)$ & \multicolumn{2}{c}{$\cdots$} &
98&.1(1)\% \\

SJ-DMC         & $-0$&$149\,177(8)$    & \multicolumn{2}{c}{$\cdots$}
& 98&.86(2)\% \\

SJ-DMC$^\ast$  & $-0$&$149\,00(5)$     & \multicolumn{2}{c}{$\cdots$}
& 98&.5(1)\%  \\

SJ-DMC$^\dagger$ & $-0$&$149\,134(9)$  & \multicolumn{2}{c}{$\cdots$}
& 98&.77(2)\% \\

SJB-DMC        & $-0$&$149\,741(2)$    & \multicolumn{2}{c}{$\cdots$}
& 100&\% \\

SJB-DMC$^\ast$ & $-0$&$149\,55(5)$     & \multicolumn{2}{c}{$\cdots$}
& 99&.6(1)\% \\

SJB-DMC$^\dagger$ & $-0$&$149\,518(9)$ & \multicolumn{2}{c}{$\cdots$}
& 99&.55(2)\% \\

\hline \hline
\end{tabular}
\caption{Energy, variance, and percentage of correlation energy
  retrieved using different methods for a 58-electron paramagnetic
  Fermi fluid of density parameter $r_s=5$ a.u.  Twist averaging has
  not been used.  ``HF,'' ``SJ-VMC,'' ``SJB-VMC,'' ``SJ-DMC,'' and
  ``SJB-DMC'' stand for Hartree-Fock theory, VMC with a Slater-Jastrow
  wave function, VMC with a Slater-Jastrow-backflow wave function, DMC
  with a Slater-Jastrow wave function, and DMC with a
  Slater-Jastrow-backflow wave function, respectively.  The DMC energy
  data have been extrapolated to zero time step. The data marked with
  an asterisk were produced by Kwon \textit{et al.},\cite{kwon} while
  the data marked with a dagger were generated by Attaccalite
  \textit{et al.}\cite{attaccalite}\ at a time step of 0.1 a.u.\
  (i.e.\ their data were not extrapolated to zero time step).  The
  fraction of the correlation energy retrieved is computed on the
  assumption that our Slater-Jastrow-backflow DMC calculation
  retrieves 100\% of the correlation
  energy. \label{table:wf_quality_rs5}}
\end{center}
\end{table}

We have optimized a three-electron term in the Jastrow factor
(together with the two-electron Jastrow terms and backflow functions)
for a paramagnetic 58-electron HEG at $r_s=5$ a.u.  The three-electron
term lowered the non-twist-averaged VMC energy from $-0.1493851(6)$ to
$-0.1495111(5)$ a.u.\ per electron.  The DMC energies at a time step
of 0.1 a.u.\ without and with the three-body Jastrow factor are
$-0.149742(2)$ and $-0.149740(2)$ a.u.\ per electron, respectively.
As expected, the inclusion of the three-body term makes an
insignificant difference to the DMC energy, because the DMC energy
depends only on the nodal surface of the trial wave function, which is
not directly affected by the Jastrow factor.  We have therefore not
used three-electron terms in our production calculations.

\subsection{Evaluating expectation values \label{sec:expval_calc}}

\subsubsection{Evaluating the MD \label{sec:md_calc}}

Let $\Psi({\bf R})$ be the trial many-electron wave function, where
  ${\bf R}=({\bf r}_1,\ldots,{\bf r}_N)$. Suppose the first
  $N_\uparrow$ electrons are spin-up and the remainder are spin-down.
  The MD of spin-up electrons can be evaluated as
\begin{equation} \rho({\bf k})=\left< \frac{1}{(2 \pi)^3} \int
  \frac{\Psi({\bf r},{\bf r}_2,\ldots,{\bf r}_N)}{\Psi({\bf R})}
  \exp[i {\bf k} \cdot ({\bf r}_1 - {\bf r})] \, d{\bf r} \right>,
  \label{eqn:mom_dist}
  \end{equation}
where the angled brackets denote an average over the set of electron
configurations generated in the VMC and DMC algorithms (which are
distributed as $|\Psi|^2$ and $|\Psi \Phi_0|$, respectively, where
$\Phi_0$ is the ground-state component of $\Psi$).  We have restricted
our attention to paramagnetic and fully ferromagnetic HEG's, so the
total MD is equal to the spin-up MD\@. The integral in the expectation
value of Eq.\ (\ref{eqn:mom_dist}) is estimated by Monte Carlo
sampling at each configuration ${\bf R}$ generated by the QMC
algorithms, and the results are averaged.  The use of a finite number
of points in the evaluation of the integral at each ${\bf R}$ does not
bias the QMC estimate of $\rho({\bf k})$.

Suppose our finite simulation cell has area $A$ and the
simulation-cell Bloch vector is ${\bf k}_s$. We may write
\begin{equation}
\frac{\Psi({\bf r},{\bf r}_2,\ldots,{\bf r}_N)}{\Psi({\bf
R})}=\frac{1}{A} \sum_{\bf G} c_{\bf G}({\bf R}) \exp[i({\bf G}+{\bf
k}_s) \cdot {\bf r}],
\end{equation}
where the $\{{\bf G}\}$ are the simulation-cell reciprocal lattice
points.  Hence it is clear that $\rho({\bf k})$ is only nonzero if
${\bf k}={\bf G}+{\bf k}_s$ for some ${\bf G}$.  The MD is only
defined for a discrete set of momenta at any given ${\bf k}_s$.  One
cannot twist average as such; instead, altering ${\bf k}_s$ leads to
the MD being defined at a different set of momenta.  We simply report
MD's obtained using ${\bf k}_s={\bf 0}$ (i.e., no twist was applied).

\subsubsection{Evaluating the SSF \label{sec:ssf_calc}}

The SSF may be evaluated as
\begin{equation} S({\bf k}) = \frac{1}{N} \left[ \left< \hat{n}({\bf
      k})\hat{n}(-{\bf k}) \right> - \left< \hat{n}({\bf k}) \right>
      \left< \hat{n}(-{\bf k}) \right> \right],
      \end{equation} where
\begin{equation} \hat{n}({\bf k})=\sum_i \exp(-i{\bf k} \cdot {\bf r}_i)
      \end{equation}
is the Fourier transform of the density operator.  $S({\bf k})$ is
only nonzero at simulation-cell ${\bf G}$ vectors, even if the
simulation-cell Bloch vector is nonzero.  We can twist average when we
calculate $S({\bf G})$.

\subsubsection{Evaluating the PCF \label{sec:pcf_calc}}

The spherically averaged PCF is
\begin{equation} g(r) = \frac{\Omega}{4 \pi r^2 N^2} \left< \sum_{i \neq j}
  \delta(|{\bf r}_i-{\bf r}_j|-r) \right>, \end{equation} which can be
evaluated by binning the electron-electron distances in the
configurations generated by the QMC algorithms.  Twist averaging
introduces no complications.

\subsubsection{Extrapolated estimation \label{sec:extrap_est}}

If $\hat{A}$ is an operator that does not commute with the Hamiltonian
then the errors in the VMC and DMC estimates $A_{\rm VMC}$ and $A_{\rm
DMC}$ of the expectation value of $\hat{A}$ are linear in the error in
the trial wave function; however, the error in the extrapolated
estimate $2A_{\rm DMC}-A_{\rm VMC}$ is quadratic in the error in the
trial wave function.\cite{foulkes_2001} We have used extrapolated
estimation in most of our calculations of expectation values.
Examples of extrapolation are shown in Figs.\
\ref{fig:HEG2D_SSF_tests} and \ref{fig:para_rs5_pcf}, and the upper
panel of Fig.\ \ref{fig:HEG2D_momdist_rs5}.  In each case the VMC,
DMC, and extrapolated estimates are in good agreement, implying that
the error resulting from the use of a DMC mixed estimate is small.
Gori-Giorgi \textit{et al.}\cite{gorigiorgi_pcf}\ used
reptation\cite{reptation} QMC to accumulate the PCF and SSF, in which
pure expectation values are obtained with respect to the fixed-node
ground-state wave function, so that extrapolation is unnecessary.

\subsection{Time-step and population-control biases \label{sec:dmc_biases}}

\begin{figure}
\begin{center}
\includegraphics[clip,scale=0.3]{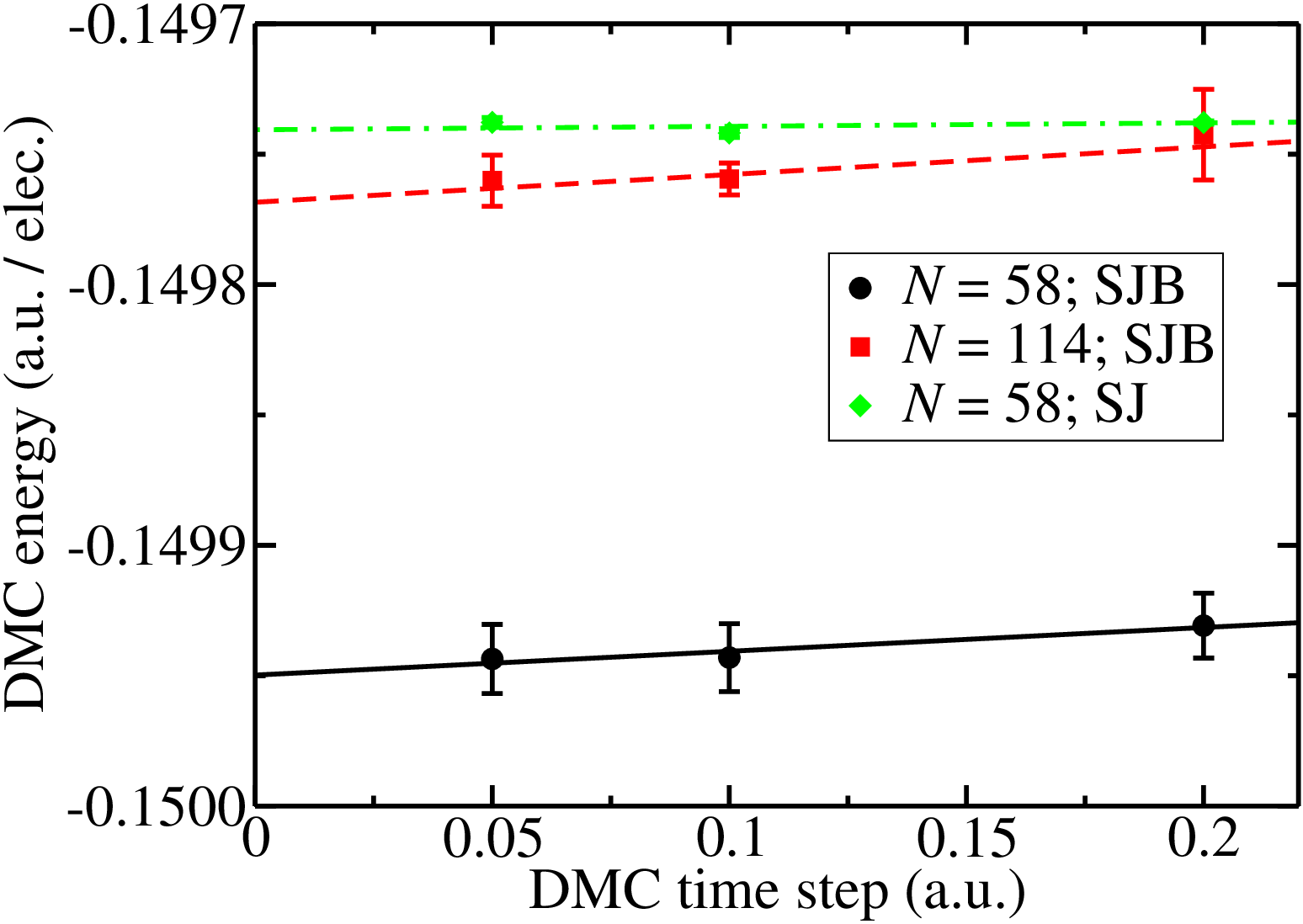}
\caption{(Color online) Twist-averaged DMC energy against time step
  for paramagnetic Fermi fluids of density parameter $r_s=5$ a.u.\ at
  different system sizes $N$.  ``SJ'' and ``SJB'' refer to
  Slater-Jastrow and Slater-Jastrow-backflow wave functions,
  respectively. \label{fig:dt_bias_example}}
\end{center}
\end{figure}

Finite-time-step errors in the twist-averaged DMC energy were removed
by linear extrapolation to zero time step. An example is shown in
Fig.\ \ref{fig:dt_bias_example}; it can be seen that the time-step
bias is in fact very small in any case. We checked that the other
expectation values were converged with respect to the time step: see
Figs.\ \ref{fig:HEG2D_momdist_rs5} and \ref{fig:HEG2D_SSF_tests}.  We
used a target population of 1600 configurations in all our DMC
calculations, making population-control bias negligible.

\subsection{Finite-size bias \label{sec:fs_bias}}

Expectation values obtained in a finite $N$-electron cell subject to
periodic boundary conditions differ from the corresponding
infinite-system values because of ``single-particle'' shell-filling
effects, as well as the neglect of long-ranged correlations and the
compression of the exchange-correlation hole into the simulation cell.
Single-particle finite-size effects can be removed from the energy,
the SSF, and the PCF by twist averaging, as explained in Sec.\
\ref{sec:trial_wfs}. We have recently demonstrated that the
finite-size error in the energy per particle in a 2D HEG falls off as
$N^{-5/4}$, enabling us accurately to extrapolate QMC energies to
infinite system size.\cite{ndd_fs} For the PCF, SSF, and MD we simply
verified that the QMC data had converged with respect to system size
(see Figs.\ \ref{fig:HEG2D_SSF_tests} and \ref{fig:para_rs5_pcf}, and
the lower panel of Fig.\ \ref{fig:HEG2D_momdist_rs5}).

\section{Results \label{sec:results}}

\subsection{Energies \label{sec:energy_results}}

DMC energies of paramagnetic Fermi fluids at different densities and
system sizes are shown in Table \ref{table:dmc_energy_data}.  Our
results for the energies of different phases of the 2D HEG at low
density are reported elsewhere.\cite{ndd_2Dheg} At $r_s=5$ and 10
a.u., Rapisarda and Senatore\cite{rapisarda} obtained infinite-system
energies of $-0.1490(1)$ and $-0.08512(2)$ a.u.\ per particle using
DMC with a Slater-Jastrow wave function.  Kwon \textit{et
al.}\cite{kwon} obtained DMC energies of $-0.2098(3)$, $-0.1495(1)$,
and $-0.08536(2)$ a.u.\ per electron at $r_s=1$, 5, and 10,
respectively, using a Slater-Jastrow-backflow wave function.  Our DMC
energies are somewhat lower than these data, as expected from the
results shown in Table \ref{table:wf_quality_rs5}.

\begin{table}
\begin{center}
\begin{tabular}{ccr@{.}l}
\hline \hline

$r_s$ (a.u.) & $N$ & \multicolumn{2}{c}{DMC energy (a.u.\ / elec.)} \\

\hline

~1         & ~50      & ~~~~~~~~~$-0$&$212\,5(2)$   \\

~1         & ~74      & $-0$&$212\,2(2)$   \\

~1         & 114      & $-0$&$211\,6(3)$   \\

~1         & $\infty$ & $-0$&$210\,4(6)$   \\

~5         & ~58      & $-0$&$149\,95(2)$  \\

~5         & 114      & $-0$&$149\,77(1)$  \\

~5         & $\infty$ & $-0$&$149\,63(3)$  \\

10         & ~58      & $-0$&$085\,504(5)$ \\

10         & ~74      & $-0$&$085\,52(1)$  \\

10         & 114      & $-0$&$085\,445(3)$ \\

10         & $\infty$ & $-0$&$085\,399(6)$ \\

\hline \hline
\end{tabular}
\caption{Twist-averaged DMC energy, extrapolated to zero time step,
  for $N$-electron paramagnetic Fermi fluids of density parameter
  $r_s$.  Where $N=\infty$, the DMC energy has been extrapolated to
  infinite system size. \label{table:dmc_energy_data}}
\end{center}
\end{table}

\begin{figure}
\begin{center}
\includegraphics[clip,scale=0.3]{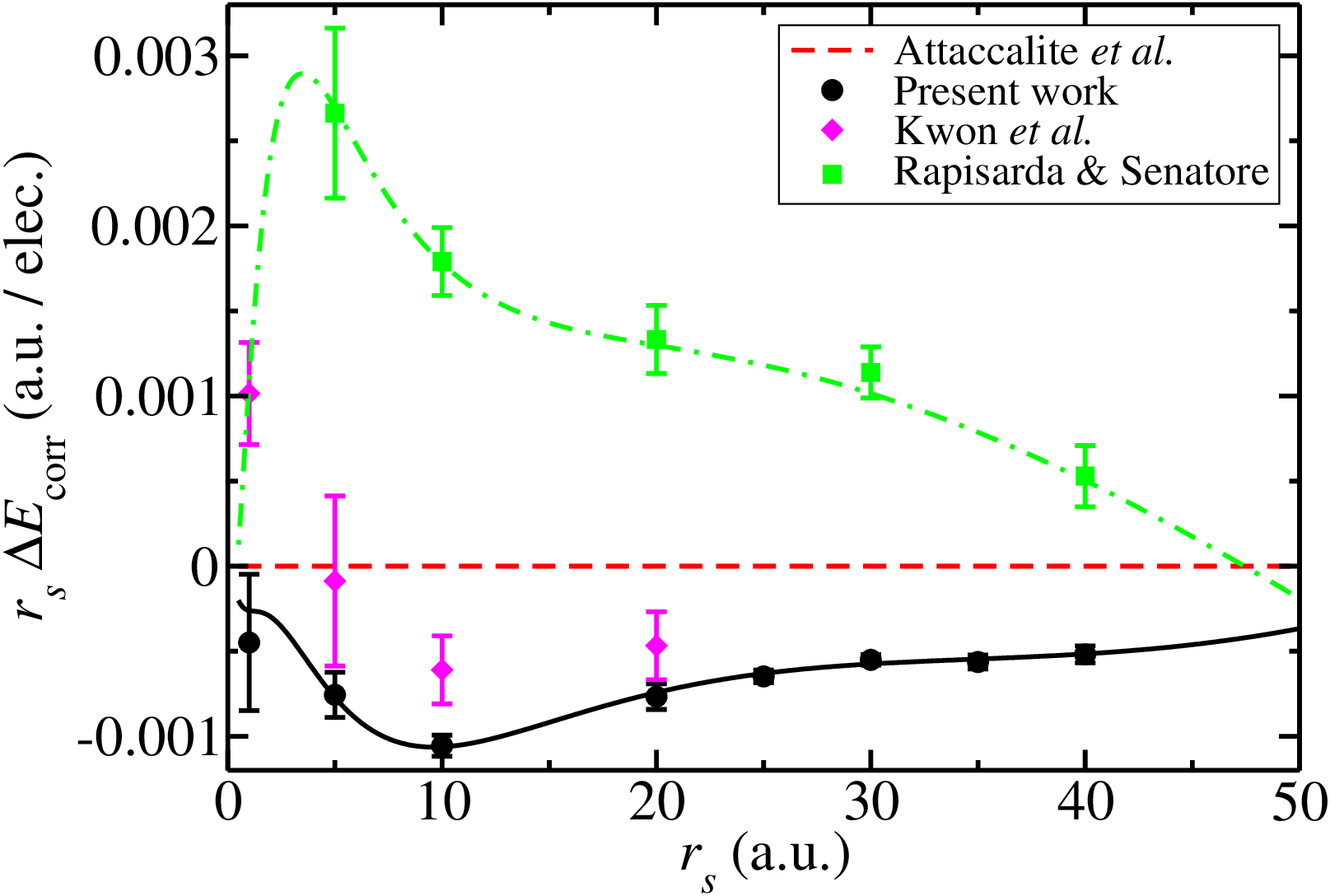}
\caption{(Color online) Correlation energy of a paramagnetic Fermi
  fluid relative to the results of Attaccalite \textit{et
  al.}\cite{attaccalite} The results obtained by Kwon \textit{et
  al.}\cite{kwon}\ and Rapisarda and Senatore\cite{rapisarda} are
  shown for comparison.  The results of Tanatar and
  Ceperley\cite{tanatar} are not shown, because they are
  systematically too low in energy.\cite{kwon} \label{fig:para_ecorr}}
\end{center}
\end{figure}

Let the correlation energy per electron $E_c$ be the difference
between the ground-state energy per electron and the Hartree-Fock
energy.  We fit the form proposed by Attaccalite \textit{et
al.}\cite{attaccalite}\ to our correlation energies for paramagnetic
HEG's:
\begin{equation} E_c = A_0+(B_0r_s+C_0r_s^2+D_0r_s^3)\log \left(
  1+\frac{1}{E_0r_s+F_0r_s^{3/2}+G_0r_s^2+H_0r_s^3} \right),
  \label{eqn:attaccalite_corr} \end{equation}
where $A_0=-0.1925$, $B_0=\sqrt{2}(10-3\pi)/(3\pi)$, and
$D_0=-A_0H_0$.  We fit to the infinite-system DMC energies shown in
Table \ref{table:dmc_energy_data} and also to the DMC energies of
low-density paramagnetic HEG's reported in Ref.\
\onlinecite{ndd_2Dheg} (at $r_s=20$, 25, 30, 35, and 40 a.u.).  Our
fitting parameters are shown in Table \ref{table:corr_fit} and the
correlation energies of paramagnetic Fermi fluids obtained by
different authors relative to that of Attaccalite \textit{et al.}\ are
shown in Fig.\ \ref{fig:para_ecorr}.  Our correlation energies are
lower than those of the other authors because of our use of flexible
backflow functions.  Equation (\ref{eqn:attaccalite_corr}) can be used
as a local-density-approximation exchange-correlation functional in
DFT calculations for 2D systems.

\begin{table}
\begin{center}
\begin{tabular}{cr@{.}l}
\hline \hline

Parameter & \multicolumn{2}{c}{Value} \\

\hline

$A_0$ & ~~~$-0$&$192\,5$    \\

$B_0$ & $0$&$086\,313\,631$ \\

$C_0$ & $0$&$069\,795\,68$  \\

$D_0$ & $0$&$0$             \\

$E_0$ & $1$&$053\,100\,3$   \\

$F_0$ & $0$&$040\,691\,22$  \\

$G_0$ & $0$&$360\,595\,3$   \\

$H_0$ & $0$&$0$             \\

\hline \hline
\end{tabular}
\caption{Fitting parameters in Eq.\ (\ref{eqn:attaccalite_corr}) for
  the correlation energy of a paramagnetic HEG\@. Equation
  (\ref{eqn:attaccalite_corr}) was fitted to the infinite-system DMC
  energies given in Table \ref{table:dmc_energy_data} and the
  low-density energy data given in Ref.\
  \onlinecite{ndd_2Dheg}. It was found that fixing $H_0=0$ did not
  affect the quality of the fit.  \label{table:corr_fit}}
\end{center}
\end{table}

Unlike Attaccalite \textit{et al.},\cite{attaccalite} we fit Eq.\
(\ref{eqn:attaccalite_corr}) to paramagnetic data only; we do not
attempt to calculate the spin-polarization-dependence of the energy of
the HEG\@.

\subsection{MD's \label{sec:md}}

\begin{figure}
\begin{center}
\includegraphics[clip,scale=0.3]{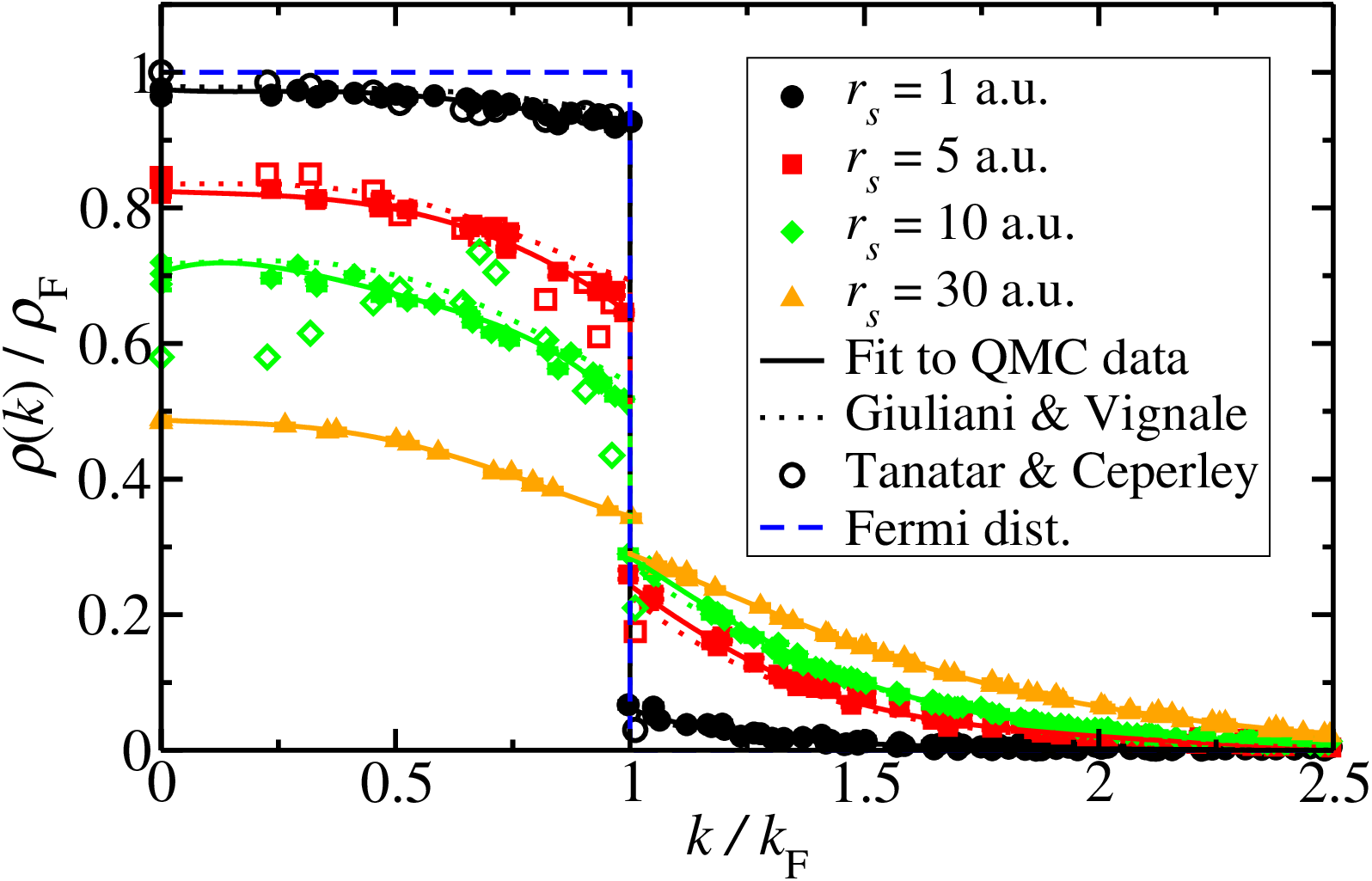}
\caption{(Color online) Extrapolated MD $\rho(k)$ for paramagnetic
Fermi fluids.  $k_{\rm F}=\sqrt{2}/r_s$ is the Fermi wave vector of
the paramagnetic fluid and $\rho_{\rm F}=r_s^2/(2\pi)$ is the value of
the Fermi distribution.  The results were obtained using a
Slater-Jastrow-backflow wave function and a variety of system sizes
with $N\geq 50$ in each case.  Twist averaging was not used.  For
comparison, we have plotted the MD obtained by Tanatar and
Ceperley\cite{tanatar} (open symbols) and Eq.\ (8.133) of Ref.\
\onlinecite{giuliani} (dotted lines). \label{fig:HEG2D_momdist}}
\end{center}
\end{figure}

\begin{figure}
\begin{center}
\includegraphics[clip,scale=0.3]{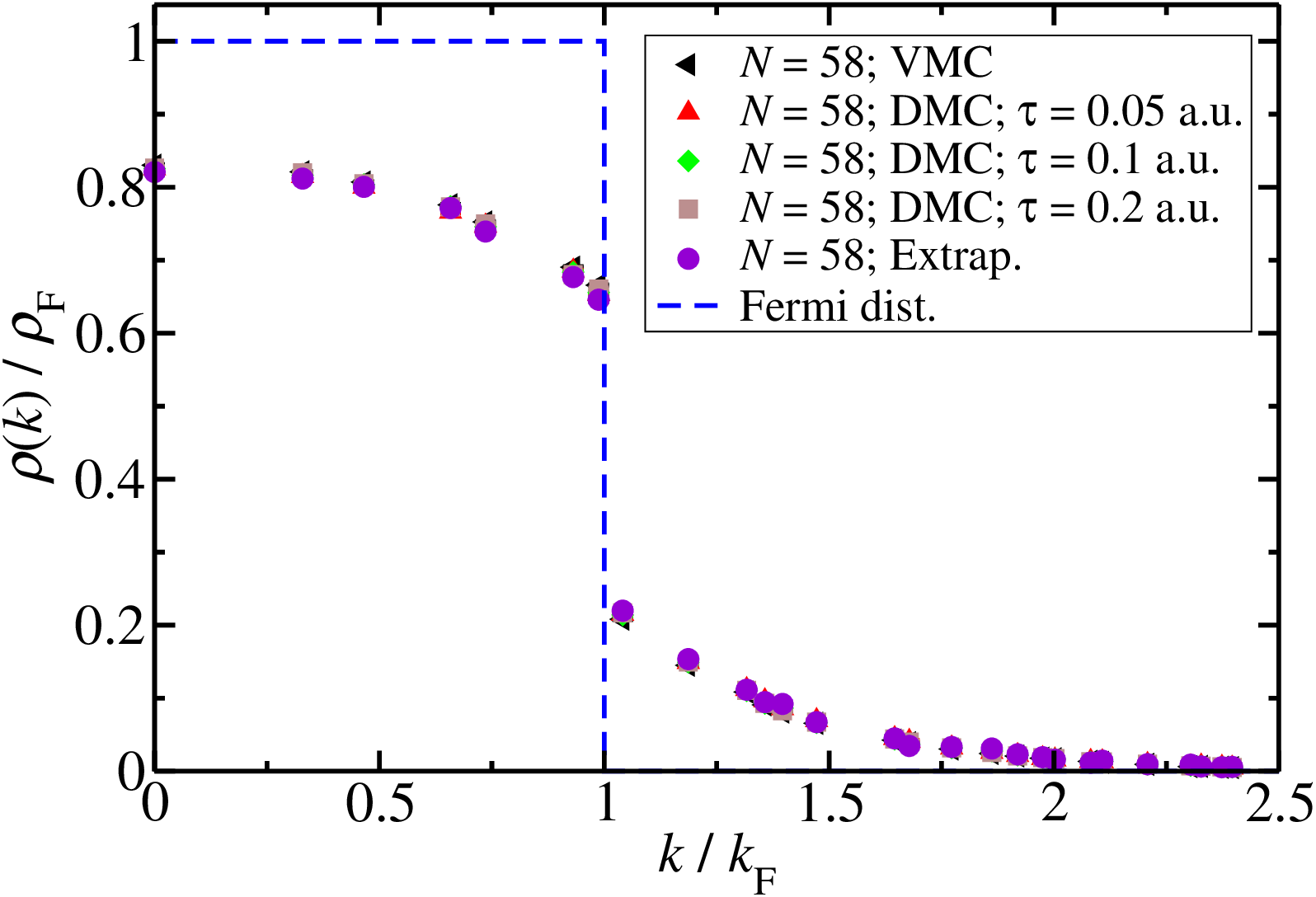} \\
\includegraphics[clip,scale=0.3]{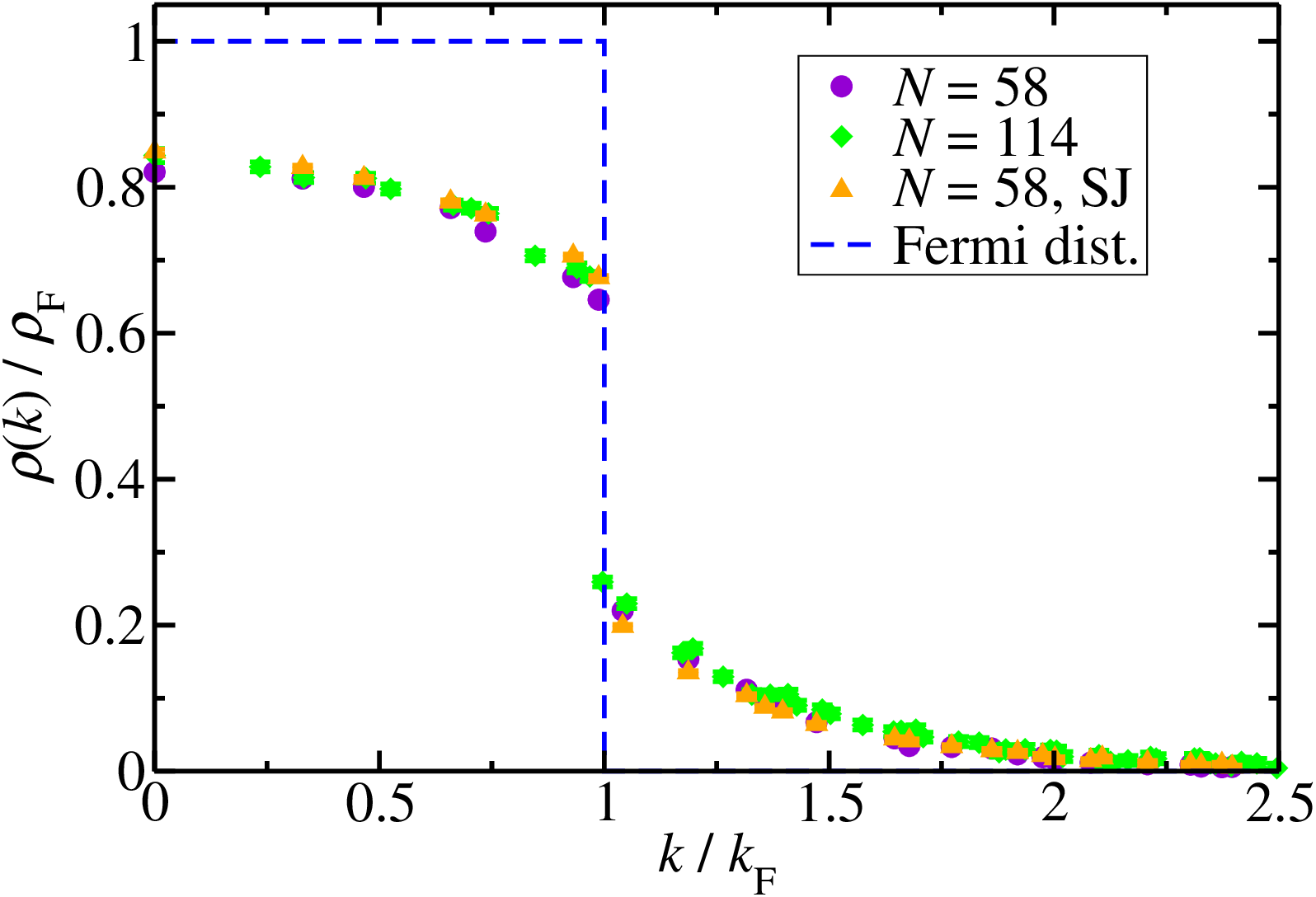}
\caption{(Color online) MD $\rho(k)$ for paramagnetic Fermi fluids at
  $r_s=5$ a.u., obtained using VMC and DMC at different system sizes
  $N$.  Backflow was used, except for the data labeled ``SJ.''  The
  upper panel shows the results obtained at $N=58$ using VMC and DMC
  with different time steps $\tau$.  The lower panel shows the effect
  of changing the system size and using backflow on the extrapolated
  MD\@.  Twist averaging was not used. $k_{\rm F}=\sqrt{2}/r_s$ is the
  Fermi wave vector of the paramagnetic fluid and $\rho_{\rm
  F}=r_s^2/(2\pi)$ is the value of the Fermi distribution.
\label{fig:HEG2D_momdist_rs5}}
\end{center}
\end{figure}

The MD's of paramagnetic HEG's are shown in Fig.\
\ref{fig:HEG2D_momdist}, and a more detailed graph of the MD at
$r_s=5$ a.u.\ is shown in Fig.\ \ref{fig:HEG2D_momdist_rs5}.  The
upper panel of Fig.\ \ref{fig:HEG2D_momdist_rs5} demonstrates that the
extrapolated estimate is accurate and that the DMC results are
converged with respect to the time step. It is clear from the lower
panel of Fig.\ \ref{fig:HEG2D_momdist_rs5} that, although backflow
makes a significant improvement to the QMC energy estimates, it has
very little effect on the MD\@. The inclusion of backflow results in a
small transfer of weight to wavevectors above the Fermi edge, as
expected, because a greater fraction of correlation energy is
retrieved.  It can also be seen that the MD's obtained at different
system sizes are in agreement.  We have therefore plotted data
obtained at different system sizes together in Fig.\
\ref{fig:HEG2D_momdist}. To our knowledge, the only previous QMC
studies of the MD of the 2D HEG are those of Tanatar and
Ceperley\cite{tanatar} and Conti.\cite{conti} At $r_s=10$ a.u.,
Tanatar and Ceperley found that the MD at small wave vectors is lower
than the value near the Fermi edge.  Our data do not show this unusual
feature.  Tanatar and Ceperley used a relatively inflexible
Slater-Jastrow wave function, which may be the reason for the
discrepancy.  Giuliani and Vignale\cite{giuliani} quote a formula for
the MD, which was obtained by fitting to QMC data generated by
Conti.\cite{conti} We have fitted our MD's to a simplified version of
the form suggested in Ref.\ \onlinecite{giuliani}:
\begin{equation}
\rho=\frac{1}{2} \left\{ \begin{array}{lr}
  a_0+a_1x+a_2x^2+a_3x^3+a_4x^4 & ~~~{\rm if}~x<\sqrt{2} \\
  \frac{4g(0)r_s^2}{x^6}+\left(a_7+a_8x+a_9x^2\right) \exp \left[
  -\frac{\left(x-\sqrt{2}\right)^2}{a_6^2} \right] & {\rm
  if}~x>\sqrt{2}
\end{array} \right., \label{eqn:md_fit} \end{equation} where $x=r_s k$
  and the $a_i$ are fitting parameters.  $g(0)$ is the contact PCF,
which we evaluated using Eq.\ (\ref{eqn:g0}).  The fitted parameters
are given in Table \ref{table:fit_MD}.  Our values for the
discontinuity at the Fermi edge are slightly smaller than those reported
in Ref.\ \onlinecite{giuliani}.

\begin{table}
\begin{center}
\begin{tabular}{c|r@{.}lr@{.}lr@{.}lr@{.}l}
\hline \hline

$r_s$ (a.u.) & \multicolumn{2}{c}{1} & \multicolumn{2}{c}{5} &
 \multicolumn{2}{c}{10} & \multicolumn{2}{c}{30} \\

\hline

$a_0$    & $1$&$950$      & $1$&$649$      & $1$&$410$     &
$0$&$974\,5$  \\

$a_1$    & $-0$&$073\,42$ & $-0$&$038\,99$ & $0$&$336\,6$  &
$-0$&$053\,5$ \\

$a_2$    & $0$&$280\,5$   & $0$&$074\,18$  & $-1$&$199$    &
$0$&$150\,9$  \\

$a_3$    & $-0$&$388\,4$  & $-0$&$192\,0$  & $1$&$148$     &
$-0$&$388\,8$ \\

$a_4$    & $0$&$136\,5$   & $0$&$021\,98$  & $-0$&$436\,3$ &
$0$&$147\,3$  \\

$a_6$    &  $1$&$171$     &  $1$&$017$     &  $1$&$035$    & $1$&$379$
\\

$a_7$    &  $0$&$164\,8$  &  $1$&$682$     &  $2$&$091$    & $1$&$428$
\\

$a_8$    & $-0$&$113\,5$  & $-1$&$282$     & $-1$&$566$    &
$-0$&$826\,4$ \\

$a_9$    &  $0$&$021\,19$ &  $0$&$277\,3$  &  $0$&$343\,8$ &
$0$&$159\,9$  \\

$Z$      &  $0$&$866$     &  $0$&$398$     &  $0$&$209$    &
$0$&$055\,5$  \\

\hline \hline
\end{tabular}
\caption{Fitting parameters in Eq.\ (\ref{eqn:md_fit}) for the MD and
  discontinuity $Z$ in the MD at the Fermi edge.  The MD is normalized
  such that the Fermi distribution is 1.
  \label{table:fit_MD}}
\end{center}
\end{table}

\subsection{SSF's \label{sec:sf}}

\begin{figure}
\begin{center}
\includegraphics[clip,scale=0.3]{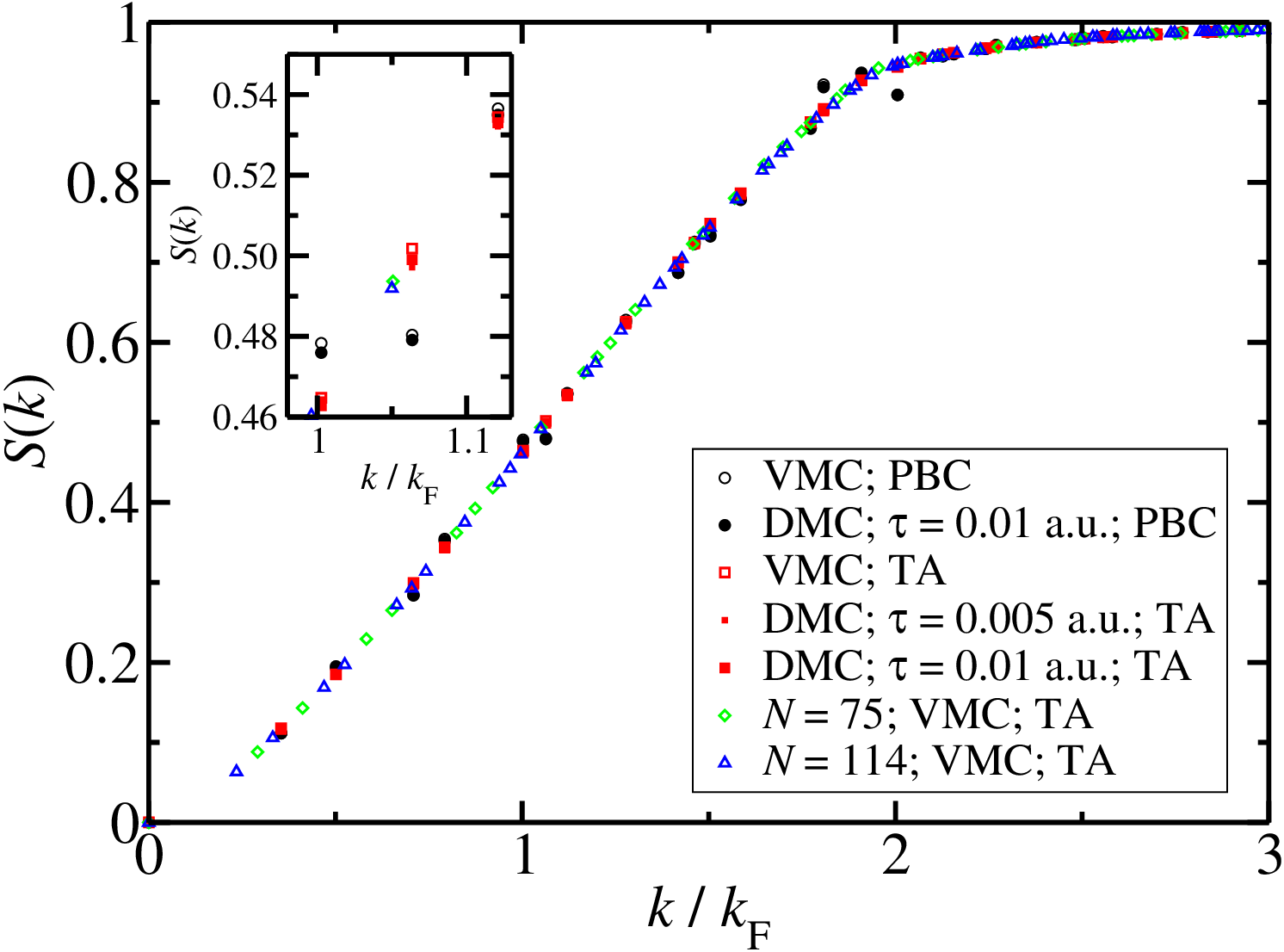}
\caption{(Color online) SSF $S(k)$ for a paramagnetic HEG of density
parameter $r_s=1$ a.u. $k_{\rm F}=\sqrt{2}/r_s$ is the Fermi wave
vector of the paramagnetic fluid. The system size was $N=50$
electrons, except where indicated otherwise. Results obtained with
${\bf k}_s={\bf 0}$ (``PBC'') and twist averaging (``TA'') are shown.
A Slater-Jastrow-backflow wave function was used in each case.  The
inset shows in greater detail one of the regions in which the
twist-averaged and non-twist-averaged data disagree.
\label{fig:HEG2D_SSF_tests}}
\end{center}
\end{figure}

VMC and DMC SSF's of a 50-electron paramagnetic HEG at $r_s=1$ a.u.\
are shown in Fig.\ \ref{fig:HEG2D_SSF_tests}.  It can be seen that the
difference between the VMC and DMC data is in most cases smaller than
the difference between the two sets of DMC data, implying that the
errors due to extrapolated estimation are small. On the other hand,
the difference between the data with ${\bf k}_s={\bf 0}$ and the
twist-averaged data is significant.  In particular, the former has
some unusual features close to integer multiples of the Fermi wave
vector, one of which is shown in the inset to Fig.\
\ref{fig:HEG2D_SSF_tests}.  Elsewhere, twist averaging has only a
small effect. At all densities we find that the twist-averaged SSF in
a 50- or 58-electron cell is in agreement with the SSF in a
114-electron cell, as can be seen in Fig.\ \ref{fig:HEG2D_SSF_tests}.
Since the statistical errors are less significant at the smaller
system sizes, we have used $N=50$ or 58 electrons in the high-density
data reported below.

\begin{figure}
\begin{center}
\includegraphics[clip,scale=0.3]{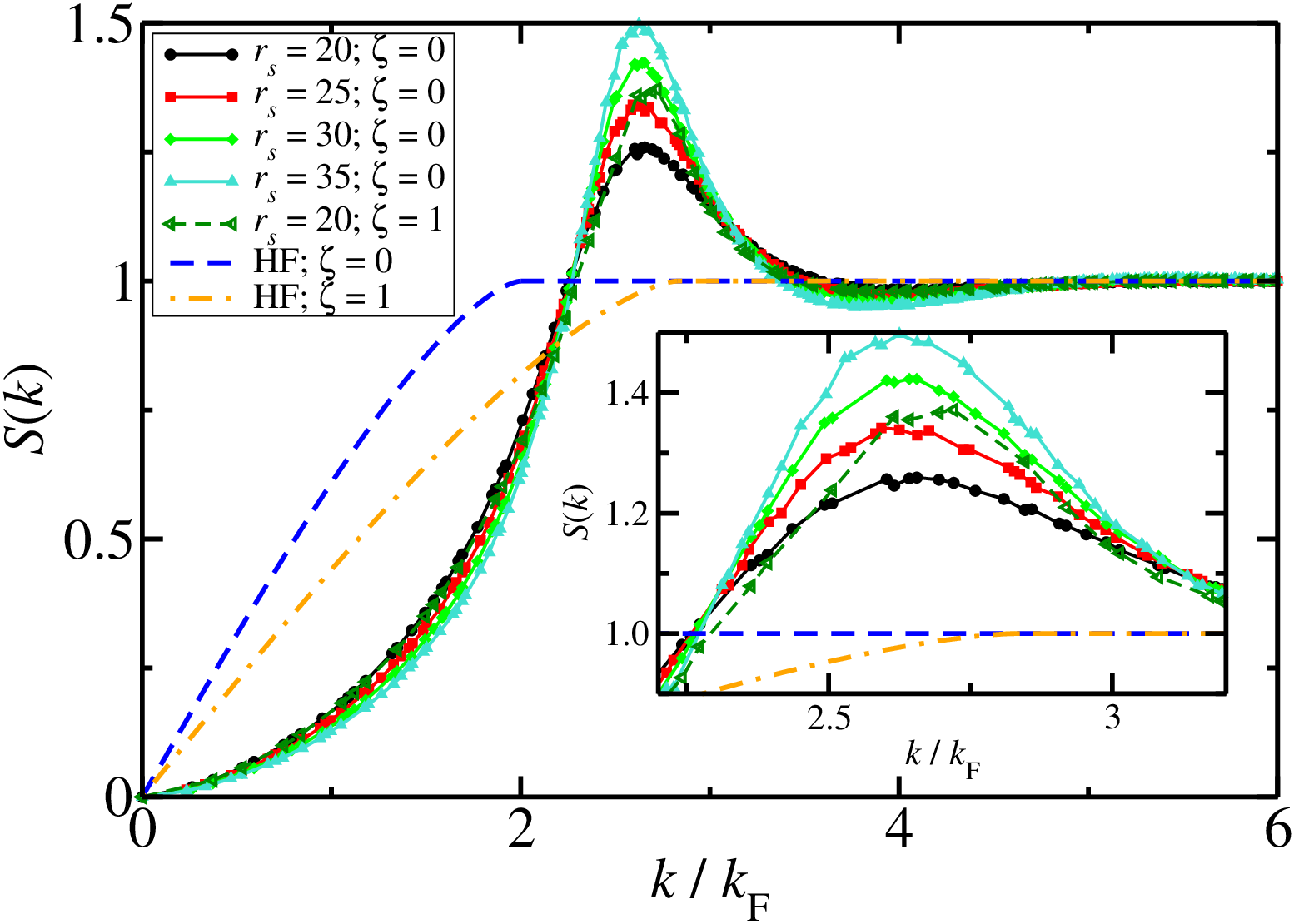}
\caption{(Color online) Extrapolated SSF $S(k)$ for HEG's of density
parameter $r_s$ and spin polarization $\zeta$.  $k_{\rm
F}=\sqrt{2}/r_s$ is the Fermi wave vector of the paramagnetic fluid.
The results were obtained using a Slater-Jastrow-backflow wave
function and twist averaging.  System sizes of $N=90$, 114, 90, and
114 were used in the paramagnetic calculations at $r_s=20$, 25, 30,
and 35 a.u., respectively, and a system size of $N=45$ was used in the
ferromagnetic calculation at $r_s=20$ a.u.  The inset shows the peak
in greater detail.
\label{fig:HEG2D_lowden_SSF}}
\end{center}
\end{figure}

\begin{figure}
\begin{center}
\includegraphics[clip,scale=0.3]{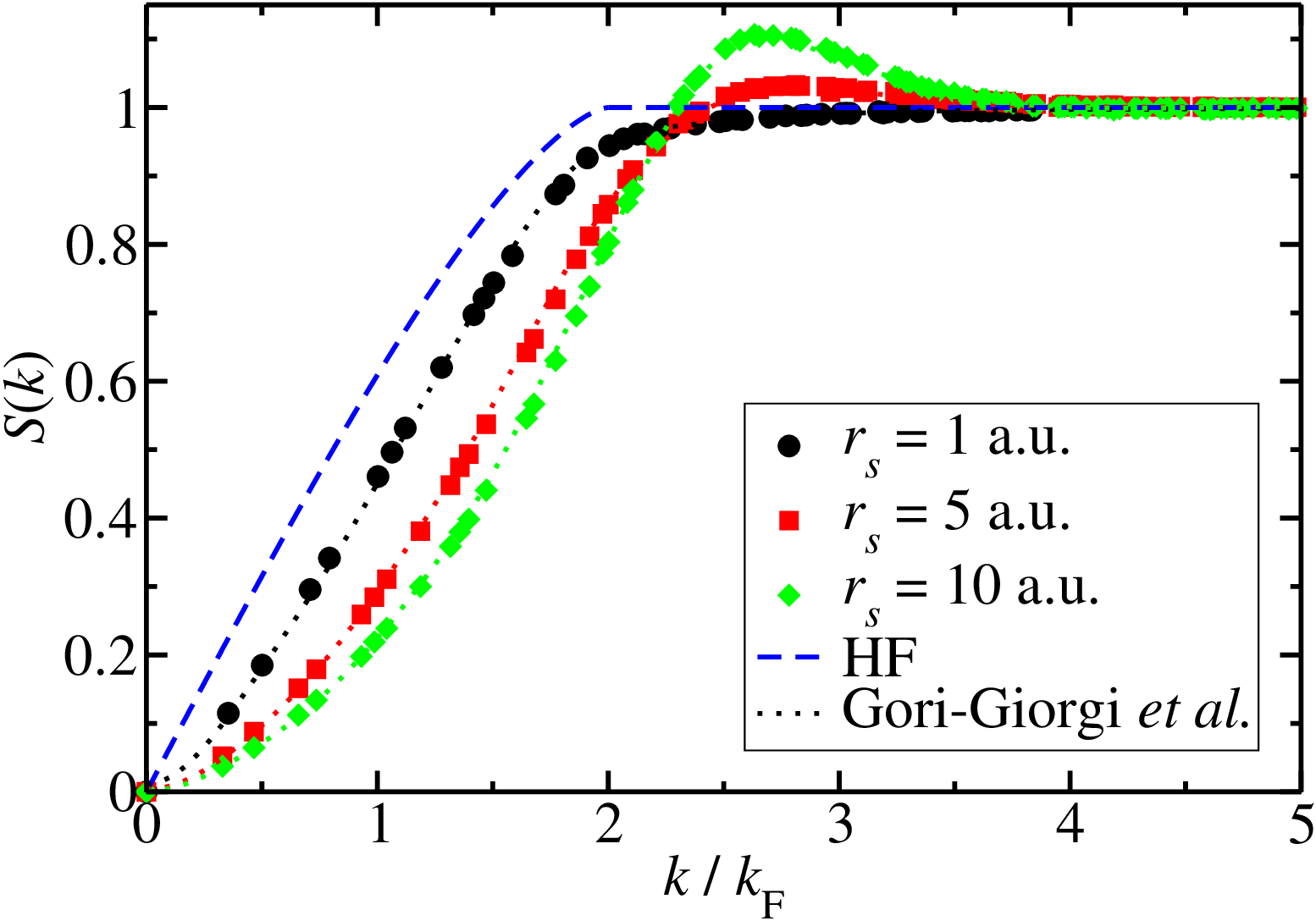}
\caption{(Color online) Extrapolated SSF $S(k)$ for paramagnetic HEG's
at high density. $k_{\rm F}=\sqrt{2}/r_s$ is the Fermi wave
vector. Slater-Jastrow-backflow wave functions and twist averaging
were used.  The system sizes are $N=50$, 58, and 58 at $r_s=1$, 5, and
10 a.u., respectively.  The curve marked ``HF'' shows the Hartree-Fock
SSF\@. The dotted lines show the SSF obtained by taking the Fourier
transform of the PCF data of Gori-Giorgi \textit{et
al.}\cite{gorigiorgi_pcf} \label{fig:para_sf_highden}}
\end{center}
\end{figure}

The SSF's of paramagnetic and ferromagnetic fluids at low density are
shown in Fig.\ \ref{fig:HEG2D_lowden_SSF}, while the SSF's of
paramagnetic fluids at high density are shown in Fig.\
\ref{fig:para_sf_highden}.  It can be seen that a peak in the SSF at
about $k=2.62 k_{\rm F}$ appears at low density, perhaps due to
incipient Wigner crystallization.  Our SSF's are in good agreement
with those of Gori-Giorgi \textit{et al.}\cite{gorigiorgi_pcf}

\subsection{PCF's \label{sec:pcf}}

\begin{figure}
\begin{center}
\includegraphics[clip,scale=0.3]{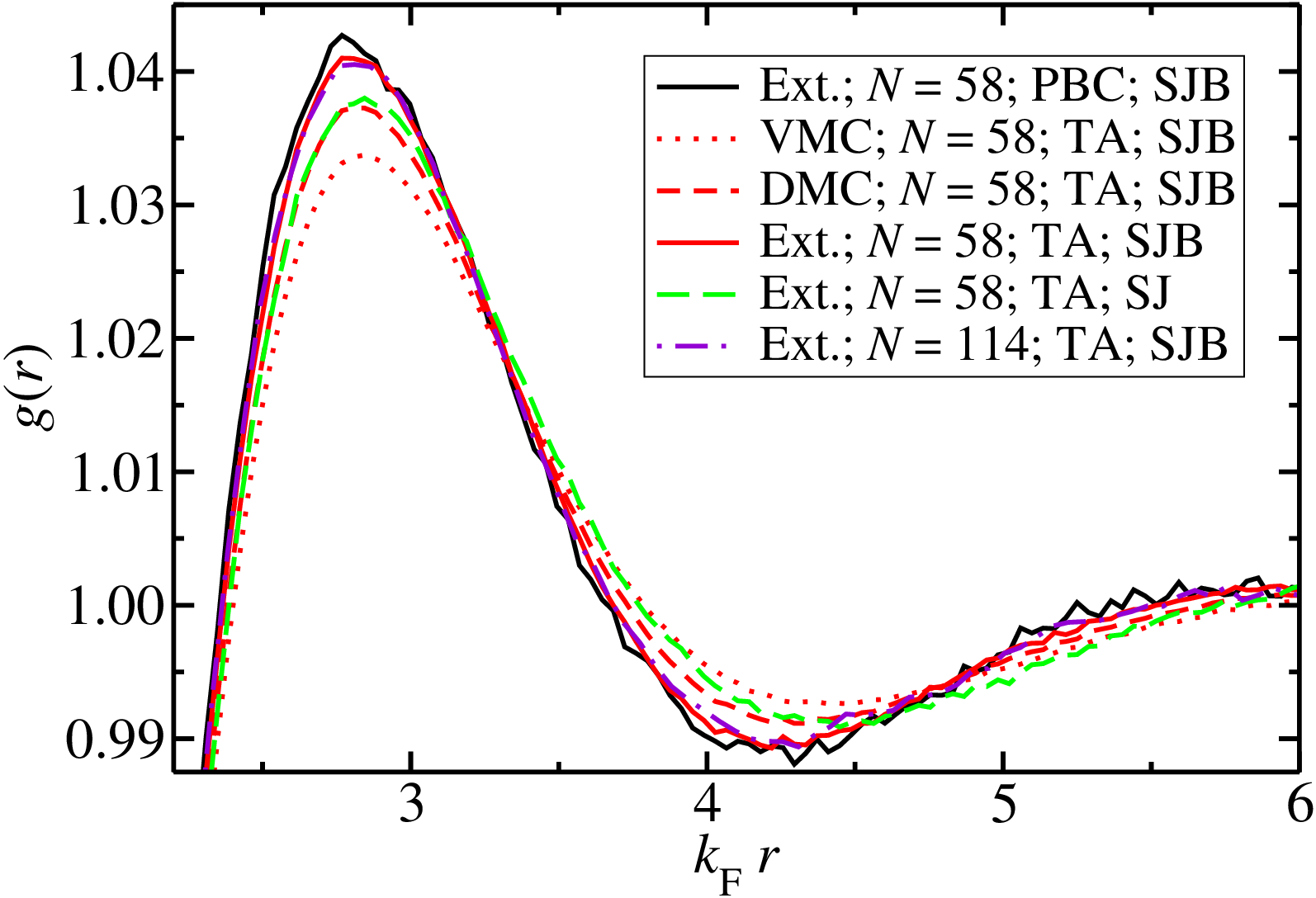}
\caption{(Color online) VMC, DMC, and extrapolated (``Ext.'') PCF's
  $g(r)$ for a paramagnetic Fermi fluid of density parameter $r_s=5$
  a.u.\ at different system sizes $N$.  Twist averaging was used in
  the curves labeled ``TA,'' but not in the one labeled ``PBC.''
  Slater-Jastrow and Slater-Jastrow-backflow wave functions were used
  in the curves labeled ``SJ'' and ``SJB,'' respectively.  $k_{\rm
  F}=\sqrt{2}/r_s$ is the Fermi wave vector of the paramagnetic fluid.
\label{fig:para_rs5_pcf}}
\end{center}
\end{figure}

We compare PCF's obtained at different system sizes using different
QMC methods for a paramagnetic HEG at $r_s=5$ a.u.\ in Fig.\
\ref{fig:para_rs5_pcf}.  The difference between the VMC and DMC PCF's,
and the difference between extrapolated PCF's obtained with and
without backflow is small, implying that the error in the extrapolated
PCF is small.  Twist averaging has a small effect on the PCF, but the
twist-averaged PCF's at $N=58$ and $N=114$ electrons are very similar,
implying that the finite-size error in the twist-averaged PCF at
$N=58$ is small. We have also verified that the PCF is converged with
respect to the DMC time step.

\begin{figure}
\begin{center}
\includegraphics[clip,scale=0.3]{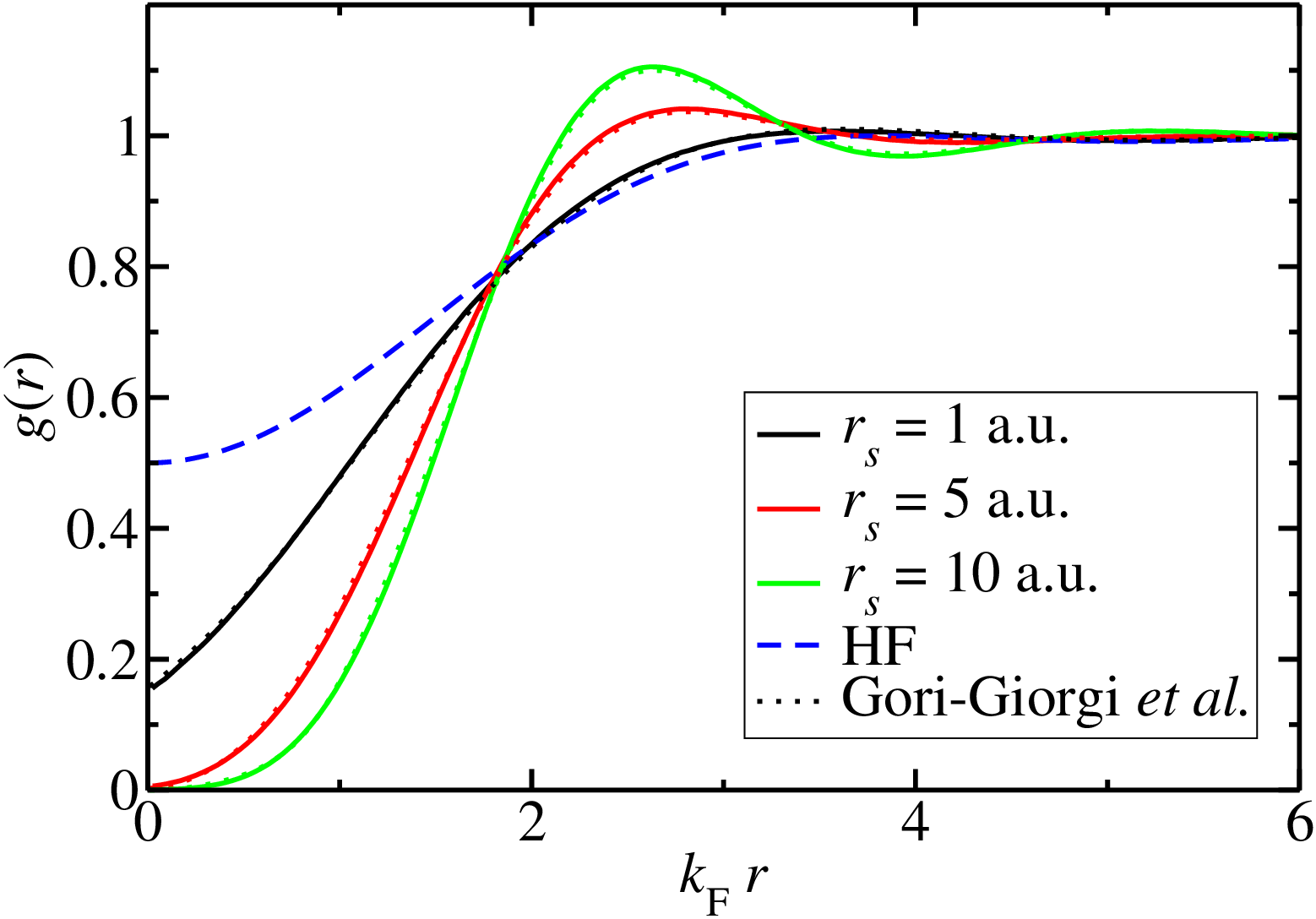}
\caption{(Color online) Extrapolated total PCF $g(r)$ for paramagnetic
  Fermi fluids of density parameter $r_s$.  $k_{\rm F}=\sqrt{2}/r_s$
  is the Fermi wave vector of the paramagnetic fluid.  ``HF'' stands
  for Hartree-Fock theory. Twist averaging was used and the QMC
  calculations were performed at system sizes of $N=50$, 58, and 58
  electrons at $r_s=1$, 5, and 10 a.u.  The dotted lines show the data
  of Gori-Giorgi \textit{et al.},\cite{gorigiorgi_pcf} which are
  almost indistinguishable from our data.
  \label{fig:para_pcf}}
\end{center}
\end{figure}

\begin{figure}
\begin{center}
\includegraphics[clip,scale=0.3]{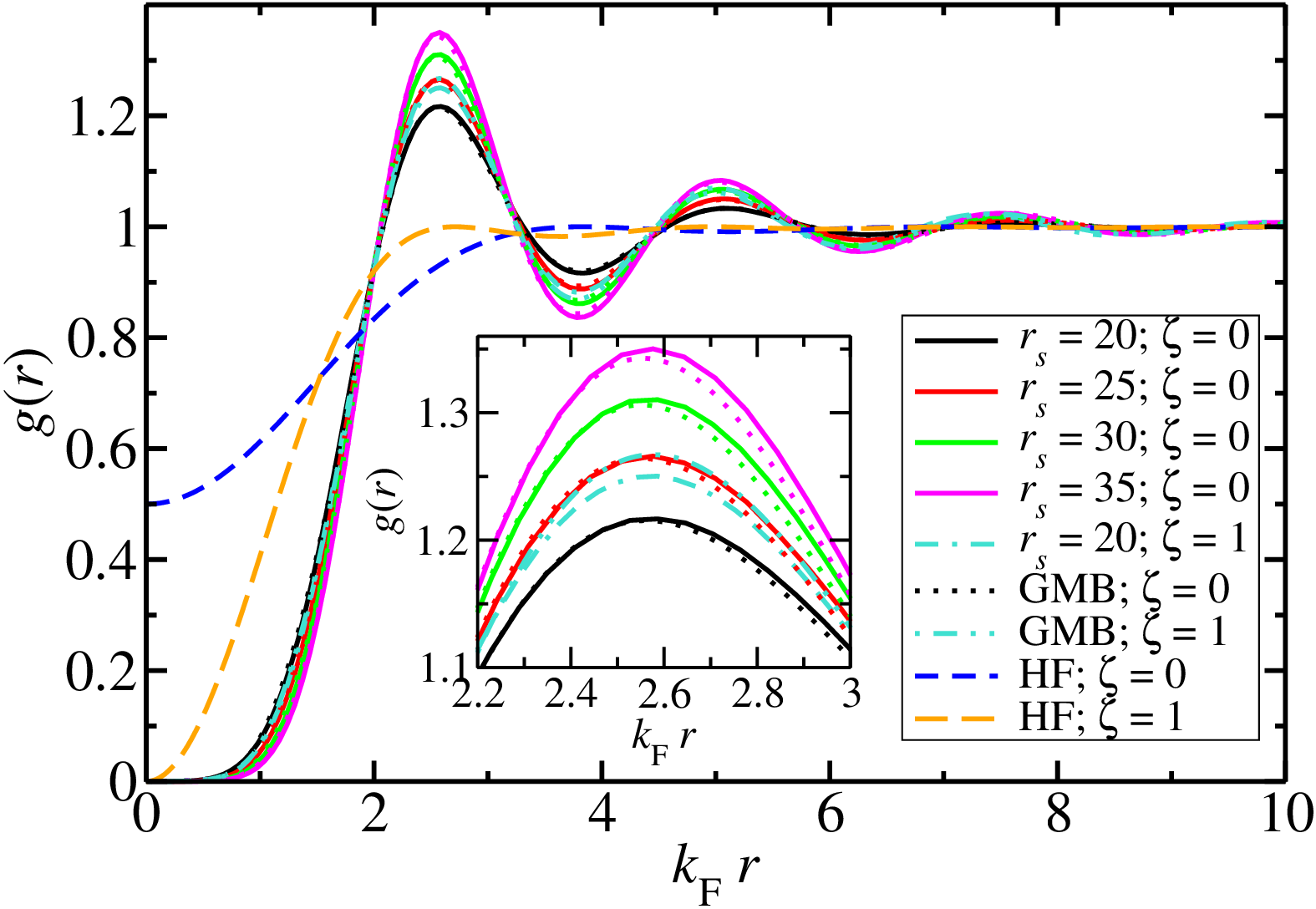}
\caption{(Color online) Extrapolated total PCF $g(r)$ for Fermi fluids
of density parameter $r_s$ and spin polarization $\zeta$.  $k_{\rm
F}=\sqrt{2}/r_s$ is the Fermi wave vector of the paramagnetic
fluid. ``GMB'' denotes the work of Gori-Giorgi \textit{et
al.}\cite{gorigiorgi_pcf}\ (shown by dotted lines), while ``HF''
stands for Hartree-Fock theory. Twist averaging was used and the QMC
calculations were performed at system sizes of $N=90$, 114, 90, and
114 at $r_s=20$, 25, 30, and 35, respectively.
\label{fig:HEG2D_PCF}}
\end{center}
\end{figure}

Our PCF's are shown in Figs.\ \ref{fig:para_pcf} (high density) and
\ref{fig:HEG2D_PCF} (low density), along with the results of
Gori-Giorgi \textit{et al.}\cite{gorigiorgi_pcf}  Our PCF's are in
good agreement with those of Gori-Giorgi \textit{et al.}\ (as
expected, from the SSF results in Fig.\ \ref{fig:para_sf_highden}).

\begin{figure}
\begin{center}
\includegraphics[clip,scale=0.3]{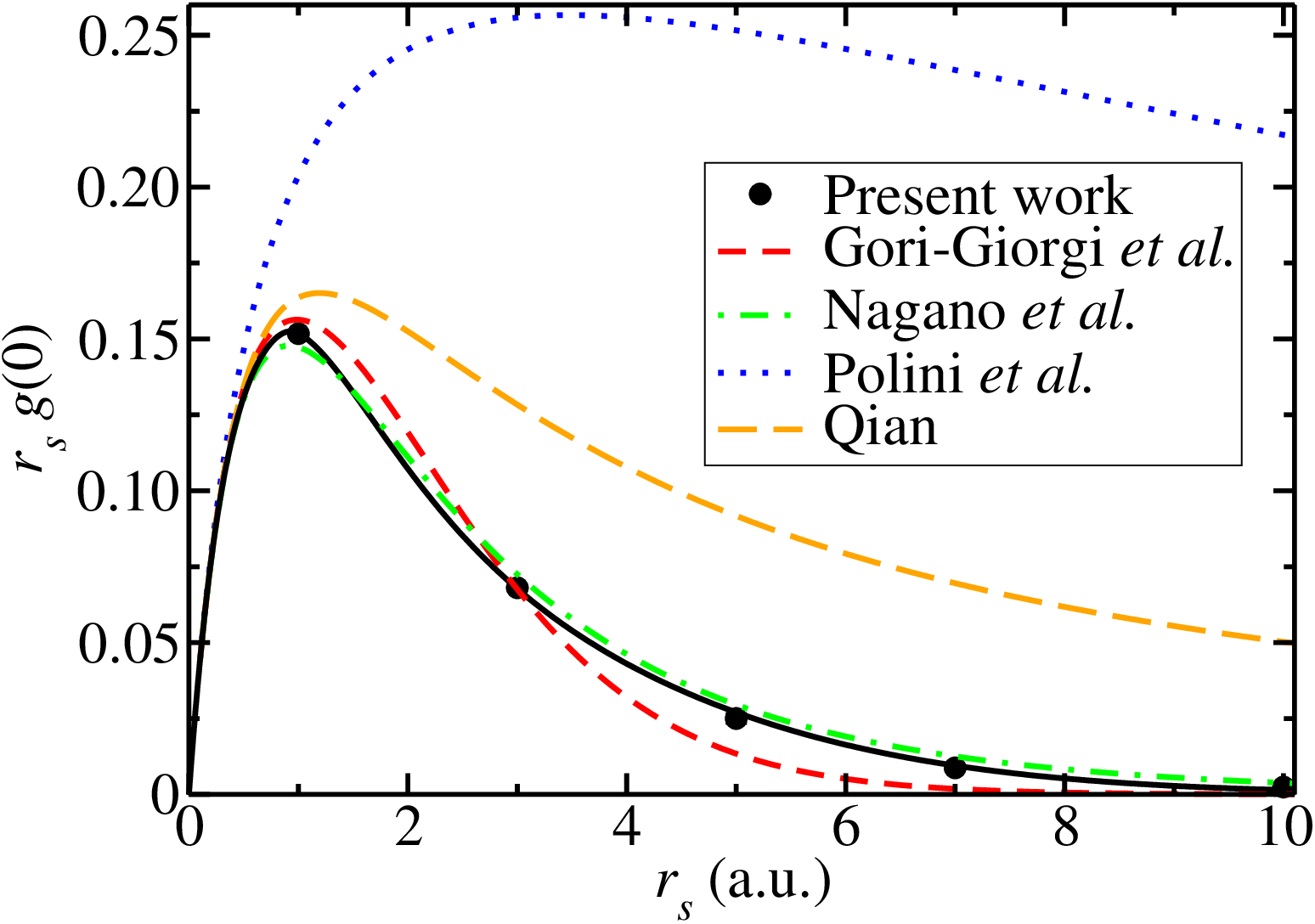}
\caption{(Color online) Contact PCF $g(0)$ against density parameter
  $r_s$, as calculated by different authors: the present work (see
  Table \ref{table:contact_pcf}), Gori-Giorgi \textit{et
  al.},\cite{gorigiorgi_pcf} Nagano \textit{et al.},\cite{nagano}
  Polini \textit{et al.},\cite{polini} and Qian.\cite{qian}
  \label{fig:contact_pcf}}
\end{center}
\end{figure}

The contact PCF $g(0)$ is especially important in the construction of
generalized-gradient-approximation exchange-correlation
functionals.\cite{giuliani} We give our $g(0)$ values in Table
\ref{table:contact_pcf}, and we plot $r_s g(0)$ against $r_s$ in Fig.\
\ref{fig:contact_pcf}, along with some other results in the
literature.  Because our PCF's have converged with respect to system
size, and our VMC and DMC results agree with each other when backflow
is used, we have simply averaged our VMC and DMC $g(0)$ data at
different system sizes in order to reduce the statistical noise in our
estimate of the contact PCF\@.  Our results are in reasonably good
agreement with the fit to the earlier QMC data of Gori-Giorgi
\textit{et al.}\cite{gorigiorgi_pcf} and also with the expression for
$g(0)$ obtained using ladder theory by Nagano \textit{et
al.}\cite{nagano} Interestingly, our results clearly disagree with the
more recent calculation of $g(0)$ within ladder theory by
Qian,\cite{qian} which involved fewer approximations than the work of
Nagano \textit{et al.}  The close agreement between our results and
those of Nagano \textit{et al.}\ must therefore be regarded as a
coincidence.  The fact that our QMC calculations, using a different
trial wave function, are consistent with the data of Gori-Giorgi
\textit{et al.}\ strongly suggests that the QMC results for $g(0)$ are
reliable, whereas ladder theory is of limited use at low
densities. Our results are also in clear disagreement with the formula
proposed by Polini \textit{et al.},\cite{polini} which interpolates
between the results of ladder theory at high-density (where it should
be exact) and a partial-wave analysis at low density.

\begin{table}
\begin{center}
\begin{tabular}{cr@{.}l}
\hline \hline

$r_s$ (a.u.) & \multicolumn{2}{c}{$g(0)$} \\

\hline

~1         & $0$&$151\,7(4)$  \\

~3         & $0$&$022\,7(2)$  \\

~5         & $0$&$005\,0(3)$  \\

~7         & $0$&$001\,25(5)$ \\

10         & $0$&$000\,24(2)$ \\

\hline \hline
\end{tabular}
\caption{Contact PCF of paramagnetic HEG's at five different
  densities.  Twist-averaged VMC and DMC results obtained at different
  system sizes were averaged to obtain these data.
  \label{table:contact_pcf}}
\end{center}
\end{table}

The fit to our $g(0)$ data shown in Fig.\ \ref{fig:contact_pcf} is
\begin{equation} g(0)=\frac{1}{2} \left\{ \begin{array}{lr} \left(
    1+Ar_s + Br_s^2 \right) \exp(-E r_s) & r_s \geq 1 \\
    1+ar_s+br_s^2+cr_s^3 & r_s<1
  \end{array} \right., \label{eqn:g0} \end{equation} where
  $A=-0.25724$, $B=0.071116$, and $E=0.98553$ were obtained by
  fitting.  Polini \textit{et al.}\cite{polini}\ have shown that
  $\lim_{r_s\rightarrow 0} g(0)=(1/2)[1-1.372r_s]$, so we have set
  $a=-1.372$ and determined $b=0.997618888$ and $c=-0.3218467056$ by
  matching the value and derivative of $g(0)$ at $r_s=1$
  a.u.\cite{footnote:g0_fit}

\section{Conclusions \label{sec:conclusions}}

We have studied the ground-state properties of the fluid phases of the
2D HEG using QMC\@. We used highly accurate trial wave functions and
dealt with finite-size effects by twist averaging. Twist averaging
removes some strange features in the SSF, but our PCF's and SSF's are
in good agreement with analytic fits to earlier QMC
data,\cite{gorigiorgi_pcf} confirming the accuracy of these formulas.
Our MD's show some qualitative differences from earlier QMC
results,\cite{tanatar} however; in particular, we do not observe an
increase in the MD as the Fermi edge is approached at low density.
Finally, we have reported DMC energy data for the high-density 2D HEG,
which we used to construct a new exchange-correlation functional for
2D DFT calculations.

\section{Acknowledgments}

We acknowledge financial support from Jesus College, Cambridge and the
UK Engineering and Physical Sciences Research Council (EPSRC)\@.
Computing resources were provided by the Cambridge High Performance
Computing Service and HPCx.


\begin{thebibliography}{99}
\bibitem{footnote:g0_fit} One can satisfy
  $g(0)=(1/2)[1-1.372r_s+O(r_s^2)]$ by imposing $E=A+1.372$ in Eq.\
  (\ref{eqn:g0}); however, we found the corresponding fits to be poor,
  because it is inappropriate to use the high-density expansion to
  determine the exponent which controls how rapidly $g(0)$ falls off
  at low density.
\end{thebibliography}
\end{document}